\documentclass[pra,onecolumn]{revtex4}
\usepackage{amssymb,amsmath,amsfonts,amsthm}
\usepackage{graphicx,color}
\usepackage{hyperref}
\usepackage{systeme}

\newcommand{\ket}[1]{{\left\vert {#1} \right\rangle}}
\newcommand{\bra}[1]{{\left\langle {#1} \right\vert}}

\newcommand{\beq}{\begin{equation}}
\newcommand{\eeq}{\end{equation}}
\newcommand{\bea}{\begin{eqnarray}}
\newcommand{\eea}{\end{eqnarray}}

\newcommand{\ug} {\!=\!}
\newcommand{\eq}{Eq.~}

\newcommand{\rref}{Ref.~}

\begin{document}

\author{Salvatore Lorenzo$^{1,2}$, Francesco Ciccarello$^{2,3}$, G. Massimo Palma$^{2,3}$}

\affiliation{$^{1}$Quantum Technology Lab, Dipartimento di Fisica, Universit$\grave{a}$  degli Studi di Milano, 20133 Milano, Italy \& INFN, Sezione di Milano, I-20133 Milano, Italy,\\
$^{2}$Dipartimento di Fisica e Chimica, Universit$\grave{a}$  degli Studi di Palermo, via Archirafi 36, I-90123 Palermo, Italy\\
$^{3}$NEST, Istituto Nanoscienze-CNR}

\begin{abstract}

It was recently shown [S. Lorenzo {\it et al.}, Sci. Rep. {\bf 7}, 42729 (2017)] that the presence of static disorder in a bosonic bath -- whose normal modes thus become all Anderson-localised -- leads to non-Markovianity in the emission of an atom weakly coupled to it (a process which in absence of disorder is fully Markovian). Here, we extend the above analysis beyond the weak-coupling regime for a finite-band bath so as to account for band edge effects. We study the interplay of these with static disorder in the emergence of non-Markovian behaviour in terms of a suitable non-Markovianity measure. 

\end{abstract}

\title{Non-Markovian dynamics from band edge effects and static disorder}
\maketitle


\maketitle

\section{Introduction}

A major topic in open quantum systems theory \cite{books} is the understanding and the characterization of quantum non-Markovian dynamics, an area of research where remarkable progress has been made in the last few years \cite{books}\cite{reviews}. This in particular has lead to a redefinition of what is Markovian in quantum mechanics, a notion which until recently was traditionally identified with an open dynamics governed by a Gorini-Kossakowski-Lindblad-Sudarshan (GKLS) master equation (master equation) \cite{books}. Recently, various definitions of quantum Markovianity have been proposed, each being typically formulated in terms of a so called non-Markovianity measure  \cite{reviews}. This is a non-negative number which is non-zero when the dynamics is non-Markovian. Such measure typically vanishes {\it in particular} for open dynamics described by a GKSL master equation, the set of zero-measure dynamics (i.e., those that are Markovian according to the considered measure) being yet strictly larger than those fulfilling a GKSL master equation with time independent decay constants. Non-Markovianity measures were used extensively in a large variety of physical scenarios to investigate occurrence of non-Markovian  behaviour as a function of the parameters entering the dynamics \cite{reviews}

Recently, through an analysis based on non-Markovianity measures, three of us showed that a quantum emitter {\it weakly} coupled to a bosonic bath (initially in its vacuum state) acquires a non-Markovian behaviour when static disorder is introduced into the bath \cite{scirep}. This was demonstrated by considering a model where a two-level system or  ``atom" is locally coupled under the rotating-wave approximation (RWA) to a large array of coupled cavities (CCA), each cavity being resonant with the atom. Since the free dynamics of the CCA is modelled through a tight-binding Hamiltonian, the CCA normal frequencies form a finite band (centered at the atom's frequency) whose width is proportional to the nearest-neighbour inter-cavity coupling rate $J$. In the weak-coupling approximation, when the atom-cavity coupling rate is much smaller than $J$, the atom in fact ``sees" the CCA as an infinite-band bosonic bath featuring a well-defined photon group velocity. Accordingly, once excited the atom decays exponentially to its ground state: a process fully described by a GKSL master equation, hence Markovian. If now static disorder is introduced into the CCA in the form of a random detuning of the frequency of each cavity, the atom decay is no longer exponential and exhibits revivals. In these conditions, an ensemble-averaged non-Markovianity measure (in \rref\cite{scirep} it was chosen the geometric measure \cite{LorenzoPP13}) grows monotonically as a function of the disorder strength, which shows that disorder causes non-Markovian behaviour.

In this work, we extend our study of occurrence of non-Markovian effects by reconsidering the model in \rref\cite{scirep}, for which however we relax the weak-coupling approximation. Accordingly, the atom is now sensitive to the finiteness of the CCA's band, in particular it can now couple in a significant way to the CCA band-edge modes for which the photon group velocity vanishes and a van Hove singularity occurs. Such band edge effects have been long studied and shown to cause a fractional atomic decay and the formation of atom-photon bound states \cite{lambro,nota-fractional}. These phenomena are traditionally recognised as a strong signatures of a non-Markovian behaviour, an assessment which -- to our knowledge -- was never made quantitative on the basis of a non-Markovianity measure. We note that both band edge effects and static disorder are alone capable of inducing non-Markovian behaviour, which makes interesting to explore their interplay when they are simultaneously present.

This paper is organised as follows. In Section \ref{model}, we introduce the model, discuss the open dynamics which the atom undergoes and show how to compute the corresponding non-Markovianity measure. In Section \ref{results}, we study the behaviour of the non-Markovianity measure as a function of disorder and coupling strength. Finally, in Section \ref{concl}, we draw our conclusions.

\section{Model and approach} \label{model}

We consider a two-level atom $S$ interacting locally with a CCA comprising an infinite number of single-mode, lossless cavities with inter-cavity coupling rate $J$. The atom is coupled to the 0th cavity under the usual rotating wave approximation with coupling rate $g$. The Hamiltonian thus reads (we set $\hbar\ug1$ throughout)
\begin{equation}
\hat H=\omega_{a} \ket{e}\bra{e}+\hat H_{f}+g\, (\hat {\sigma}_+\hat{a}_{0}+\hat {\sigma}_-\hat{a}^\dag_{0})\,,\label{H}
\end{equation}
where the free Hamiltonian of the CCA is \cite{nota-BCs} 
\begin{eqnarray}
\hat{H}_{f}=\sum_{n=-N}^{N}\!\left[\varepsilon_n \hat{a}_n^\dag \hat{a}_n-J(\hat{a}_{n}\hat{a}_{n+1}^\dag+\hat{a}_{n}^\dag \hat{a}_{n+1})\right]\label{Hf}
\end{eqnarray}
(we are ultimately interested in the limit $N{\rightarrow}\infty$).
Here, $\hat \sigma_+=\hat \sigma_-^\dag =\ket{e}\bra{g}$ are the pseudo-spin ladder operators of $S$, whose ground and excited states $|g\rangle$ and $|e\rangle$, respectively, are separated by an energy gap $\omega_a$, while $\hat a_n$ ($\hat a_n^\dag$) annihilates (creates) a photon in the $n$th cavity whose frequency is $\varepsilon_n$. 

In absence of disorder, the cavity frequency is uniform throughout the array, i.e., $\varepsilon_n=\omega_0$ for all $n$. When $\varepsilon_n=\omega_0$  (no disorder), the CCA normal modes are plane waves indexed by $-\pi\le k\le\pi$ (for $N\rightarrow \infty$) with associated dispersion law $\omega_k=\omega_0-2J \cos k$. The single-photon spectrum of the CCA therefore consists of a finite band of width 4$J$ centered at $\omega=\omega_0$. Band edge modes thus occur for $\omega=\omega_0\pm 2J$. At these frequencies, the photon group velocity $\upsilon_k={\rm d}\omega/{\rm d}k$ vanishes while the CCA density of states accordingly diverges (van Hove singularity) \cite{lambro}.

To add static disorder in the CCA we introduce a random detuning $\delta_n$ on each cavity such that $\varepsilon_n=\omega_0+\delta_n$, where $\{\delta_n\}$ are a set of identically and independently distributed random variables according to a given probability distribution function.

The open dynamics of concern here is the atom's spontaneous emission that occurs when the CCA is initially in its vacuum state $|{\rm vac}\rangle$. Accordingly, if the atom is initially in $|g\rangle$ it will remain in this state indefinitely since $\hat H|g\rangle|{\rm vac}\rangle=0$. If the atom starts in $|e\rangle$ instead, by using that the total number of excitations is conserved the time-evolved atom-field state takes the form 
\begin{equation}
\ket{\Psi(t)}=e^{-i\hat{H}t}\ket{\Psi(0)}=\alpha(t)\ket{e}\ket{\rm vac}+\ket{g}\ket{\psi_1(t)}\label{psit}
\end{equation}
with $|\Psi(0)\rangle=|e\rangle|{\rm vac}\rangle$ and $|\psi_1(t)\rangle$ a time-dependent (unnormalised) single-photon state of the CCA. In the general case in which the atom is initially in an {\it arbitrary} state $\rho(0)$ with density matrix matrix elements $\rho_{jk} =\langle j|\rho(0)|k\rangle$ (with $j,k=g,e$), its evolved state reads \cite{books}
\begin{equation}
{\rho}(t)=
\begin{pmatrix}|\alpha(t)|^2\rho_{ee}&\alpha(t)\rho_{eg}\\
\alpha(t)^*\rho_{ge}&(1-|\alpha(t)|^2)\rho_{ee}+\rho_{gg}\end{pmatrix}\,.
\label{TLSmap}
\end{equation}

\eq(\ref{TLSmap}) defines the dynamical map  of the spontaneous emission process. For any fixed pattern of detunings $\{\delta_n\}$, namely a specific realisation of noise, 
the degree of non-Markovianity of the dynamical map (\ref{TLSmap}) can be computed through one of the non-Markovianity measures proposed in the literature. While these measures are generally inequivalent, in the case of \eq(\ref{TLSmap}) -- which is a so called amplitude damping channel \cite{NC} -- the existence of times such that ${\rm d} |\alpha(t)|/{\rm d}t{>}0$ is a necessary and sufficient condition in order for the RHP \cite{RHP}, BLP \cite{BLP} and geometric \cite{LorenzoPP13} non-Markovianity measure to be non-zero \cite{reviews}. Out of these, in line with other studies addressing atom-photon interaction dynamics \cite{tommy,usa}, we select the geometric measure  for its ease of computation and to make straighforward a comparison with the results of \rref\cite{scirep}. 

 The geometric measure definition is based on the time changes of $V(t)$, i.e., the volume of accessible states of $S$ [in our case $S$ is a two-level system and $V(0)$ is the volume of the Bloch sphere]. This volume can only decrease with time in particular for dynamics governed by a GKSL master equation, hence ${\rm d}V/{\rm d}t{>}0$ can be considered as a signature of non-Markovian behaviour. Based on this, the geometric measure is defined as $\mathcal{N}_V={1}/{V(0)}\int_{\partial_t{V(t)}>0}\!{\rm d}t\,\,\partial_t{V(t)}$, the integral being over time domains such that $V(t)$ grows. For a dynamical map of the form \eqref{TLSmap}, it turns out that \cite{LorenzoPP13}
\begin{equation}
\mathcal{N}_V=\int_{\partial_t{|\alpha(t)|}>0}\!\!{\rm d}t\,\,\frac{\rm d}{{\rm d}t}|\alpha(t)|^4\,.\label{NV}
\end{equation}
As anticipated, there are times at which $\mathcal{N}_V{>}0$ iff $|\alpha(t)|$ grows. To avoid divergences of $\cal N$, which typically occur when $|\alpha(t)|$ exhibits stationary oscillations, here in line with \rref \cite{scirep} we will use a rescaled version of the geometric measure defined by 
\begin{equation}
\mathcal{N}=\frac{\mathcal N_V}{\left|\int_{\partial_t{|\alpha(t)|}<0}\!{\rm d}t\,\partial_t|\alpha(t)|^4\right|}\label{N}
\end{equation}
(note that the integral on the denominator is over intervals on which $|\alpha(t)|$ {\it decreases}). It is easily shown \cite{scirep} that $0{\le}{\cal N}{\le}1$, hence in particular ${\cal N}=1$ indicates an extremely non-Markovian dynamics (such as vacuum Rabi oscillations of an atom in a lossless cavity)\cite{nota-rescaled}.

Clearly, each particular pattern of detunings $\{\delta_n\}$, i.e., a specific realization of static disorder, leads to a different function $\alpha(t)$, hence a different dynamical map (\ref{TLSmap}) with its own associated amount of non-Markovian behaviour as measured by ${\cal N}$. 
Note that all non-Markovianity measures are non linear function of the density operator, therefore the average non-Markovianity is not the same of the non-Markovianity of 
the averaged density operators. Furthermore all non-Markovianity measures characterise the dynamical map describing the reduced time evolution of system. 
We stress again that in our scenario each specific realization of static disorder, leads to a different dynamical map, each with a different associated amount of non-Markovianity.
Therefore, for static disorder, the correct quantity to evaluate is the {\em ensemble-average} non-Markovianity measure, which we call $\bar {\cal N}$. 
To make clear this point let us note that,  for instance, in the weak-coupling regime studied in \rref\cite{scirep} the ensemble-averaged open dynamics of the atom consists of a monotonic decay towards an asymptotic state, which yields a zero NM measure. This contrasts the fact that, in the vast majority of realizations of disorder, the atom instead undergoes revivals which are a signature of NM behavior [see \eq\eqref{NV}]. This NM behavior is instead captured by $\bar {\cal N}$, which we will thus use throughout (in line with \rref\cite{scirep}) to measure NM effects.
Nota that in the different physical scenario studied in  \rref\cite{lattice} disorder is not static, but it is instead a time dependent classical stochastic noise and therefore, in the appropriate regime, it is physically meaningfull to evaluate the non-Narkovianity of the average dynamical map)
In each single realization, we numerically track the dynamics up to time $t=T$, where $T$ is long enough for the atom to fully release the amount of initial excitation that does not overlap atom-photon bound states in the absence of disorder (i.e., for $\sigma=0$) \cite{nota-T}. We made sure that $N$ (namely the CCA's length in fact) is large enough that no emitted photon reaches its edges within the time $T$.

Each particular realisation of noise was obtained by randomly generating a detuning for each cavity according to the Gaussian probability distribution function $p(\delta)=\exp{\left[{-\frac{\delta^2}{2 \sigma ^2}}\right]}/(\sqrt{2 \pi } \sigma)$, with the standard deviation $\sigma$ thus quantifying the {\it disorder strength}. 

Throughout we set $\omega_a=\omega_0$, thereby the atomic frequency lies right at the center of the CCA band (see above).

\section{Behaviour of non-Markovianity measure}\label{results}

Figure 1(a) shows the average non-Markovianity measure against the disorder strength $\sigma$ and atom-photon coupling rate $g$.
\begin{figure}[h!]
	\begin{center}
	\includegraphics[width=\textwidth]{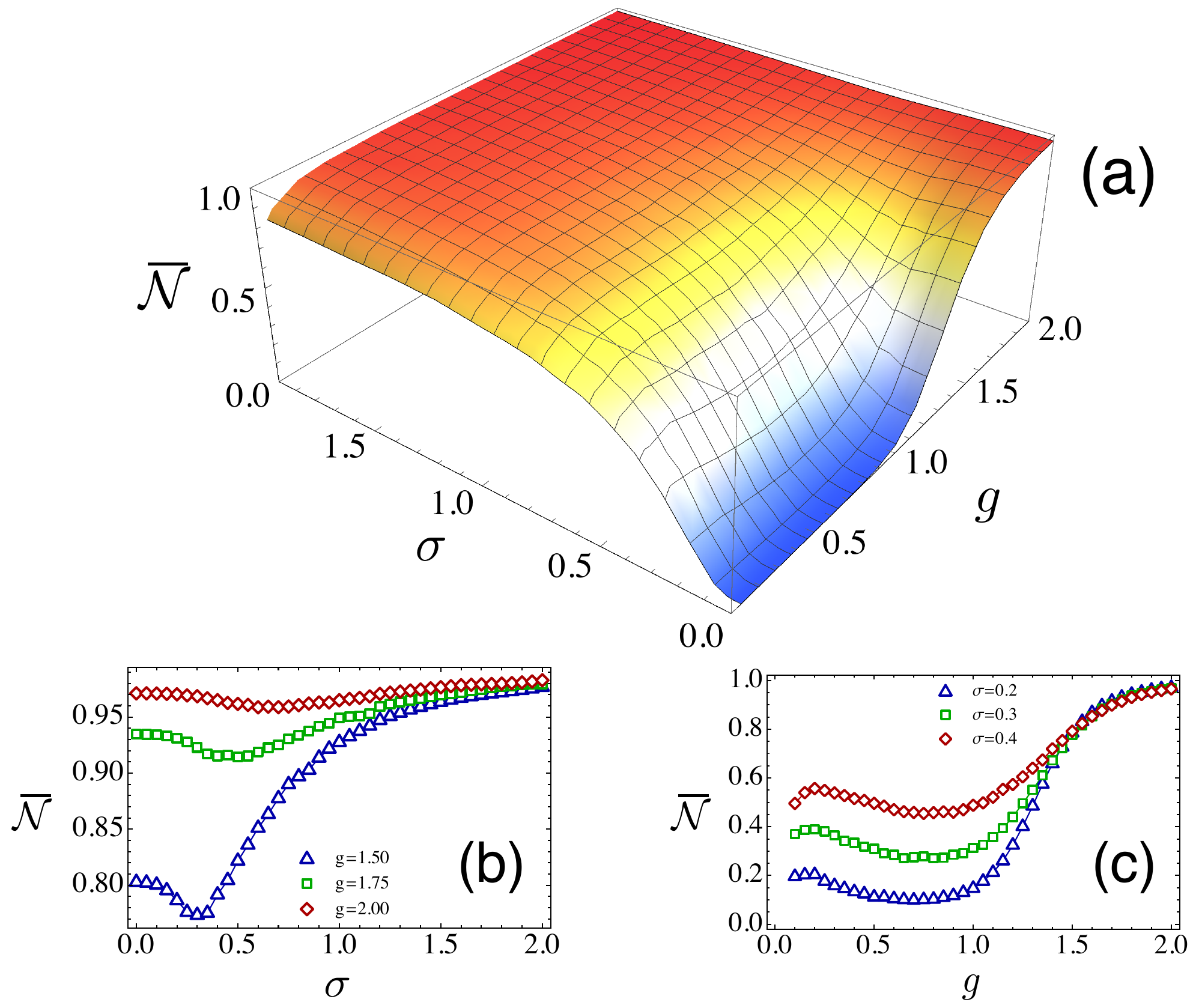}
	\caption{(a) Ensemble-averaged non-Markovianity measure $\overline{\mathcal N}$ versus $\sigma$ and $g$ (in units of $J$). In figure (b) [(c)] we plot $\overline{\mathcal N}$ versus $\sigma$ ($g$) for few representative values of $g$ ($\sigma$). Throughout, we considered a CCA made out of $N=1000$ cavities. For each value of $\sigma$, averages were performed over $4{\times}10^3$ different realizations of disorder.}
	\label{NMgauss}
	\end{center}
\end{figure}
For very small $g$, say $g\lesssim0.1J$, we retrieve the results of \rref\cite{scirep}: for increasing disorder the amount of non-Markovian behaviour monotonically grows from zero until, for large enough $\sigma$, it saturates to the maximum value $\bar {\cal N}=1$. This occurs because in the absence of disorder ($\sigma=0$) the only involved modes of the CCA are those at the centoer of the band. These modes are unbound, i.e., they are extended over the entire CCA, and have group velocity $\upsilon= 2J$ (in either direction). Accordingly, the energy released by the atom will flow away indefinitely along the CCA without ever being reabsorbed. The excited-state probability thus exhibits a monotonic decrease, leading to a non-Markovianity measure equal to zero [see \eq(\ref{NV})]. For non-zero $\sigma$, instead, the modes of the disordered CCA are all localised due to Anderson localisation. The atom will in particular strongly couple to those of frequency $\omega\simeq \omega_0$ that are most localised around the cavity $n=0$. These modes behave as an effective cavity which continuously exchanges energy with the atom. The atomic population will thus undergo revivals causing the non-Markovianity measure to be non-zero [see \eq(\ref{NV})]. For growing $\sigma$, the localisation gets stronger and stronger and the atomic revivals become more pronounced so that $\bar {\cal N}$ increases. Such increase continues until $\bar {\cal N}$ saturates to its maximum value (this happens when the disorder is so high that each mode is localised on a single cavity, giving rise to an effective Jaynes-Cummings dynamics and thus vacuum Rabi oscillations of the atom).

As $g$ is made larger, i.e., beyond the weak-coupling regime, the behaviour of $\bar{\cal N}$ has the following main features. For fixed $g$, the non-Markovianity measure {\it generally} grows with $\sigma$ until it saturates to its maximum value (we discuss this behaviour in more detail below). Yet, unlike the weak-coupling regime, $\bar{\cal N}$ generally takes a finite value even for zero disorder [see figure 1(a)], with this value becoming larger and larger as $g$ grows up. As a consequence, the overall rise of $\bar{\cal N}$ with $\sigma$ becomes less and less steep when $g$ increases until, for $g\gtrsim1.5 J$, $\bar{\cal N}$ basically takes its maximum value regardless of $\sigma$. 

Note that [see figure 1(a)], although finite, the zero-disorder value of the non-Markovianity measure remains rather small as long as $g\lesssim J$ (beyond this value it starts growing more rapidly until it saturates to $\bar{\cal N}=1$ at $g\simeq 2J$). Such zero-disorder non-Markovianity stems from the presence of a pair of atom-photon bound states whose energies lie symmetrically out of the continuum (i.e., out of the band $\omega_0-2J\le\omega\le\omega_0+2J$) \cite{Lombardo2013,calajo}. Since the state $|e\rangle|{\rm vac}\rangle$ overlaps this pair of states, a fraction of the atomic excitation undergoes stationary oscillations \cite{Lombardo2013,calajo} (hence periodic revivals) giving rise to a finite non-Markovianity measure. This effect however is significant only provided that $g\gtrsim J$ since the pair of bound states are due to the coupling of the atom to zero-velocity band-edge modes of the CCA (recall that the atom frequency lies at the center of the band whose width is ${\sim }J$). The higher $g$ the more localized around $n\ug 0$ is each atom-photon bound state until for $g$ large enough the two bound states reduce to the pair of single-excitation dressed states arising from the Jaynes-Cummings-like coupling between the atom and the cavity $n=0$.

The previous arguments in particular explain why $\bar{\cal N}{\rightarrow}1$ both for large enough $\sigma$ when $g\simeq0$ and for large enough $g$ when $\sigma=0$. Figure 1(a) yet shows that the former and latter limits hold even for arbitrary $g$ and $\sigma$, respectively. Indeed, with very strong disorder only a localised mode of the CCA overlaps $n=0$ and so the atom undergoes vacuum Rabi oscillations at a Rabi frequency depending on $g$. Regardless of the value of the Rabi frequency as well as of any detuning between the localised mode and the atom, this will exhibit stationary oscillations (at all times) that -- irrespective of their amplitude or frequency -- yield maximum non-Markovianity measure. On the other hand, for large enough $g$, the coupling between the atom and the $n=0$ cavity dominates the dynamics (no matter what the pattern of frequencies $\{\delta_n\}$ is), causing again stationary oscillations of the atom and hence maximum non-Markovianity measure. 

A careful inspection of figure 1(a) shows that the increase of the non-Markovianity measure with $\sigma$ for fixed $g$ or with $g$ with fixed $\sigma$ is not always strictly monotonic. Indeed, there is a range of intermediate values of $\sigma$, approximately between $\sigma\simeq0.2 J$ and $\sigma\simeq0.5J$, where for growing $g$ the non-Markovianity measure first increases then slightly decreases and eventually converges to its maximum value. Three specific instances of this behaviour are displayed in figure 1(c). 

 A similar non-monotonic behaviour but as a function of $\sigma$ takes place within a narrow range of values of $g$ between $g\simeq1.5J$ and $g\simeq2J$, as shown by figure 1(a) and the three numerical instances in figure 1(b). In some respects, such non-monotonic behaviour can be expected. For instance, for coupling strengths large enough to yield non-Markovianity due to band edge effects it is reasonable that the addition of disorder can cause decoherence spoiling the formation of atom-photon bound states (which are entangled \cite{Lombardo2013}). In the light of this, it is remarkable that this non-monotonicity is such a small effect [barely visible with the plot scale used in figure 1(a)], a phenomenon calling for a better understanding.

\section*{\bf Conclusions}\label{concl}

In this work, we studied the occurrence of non-Markovian behaviour, as characterized by the geometric volume non-Markovianity measure, in the open dynamics dynamics of a two-level atom emitting into a disordered CCA for arbitrary atom-photon coupling strengths. Unlike the weak-coupling regime \cite{scirep}, non-Markovian behaviour generally takes place even for zero disorder due to band edge effects. For fixed coupling strength (disorder strength), a growth of static disorder (coupling strength) causes an overall rise of the amount of non-Markovianity until this eventually saturates to its maximum value. 

\section{Acknowledgments} \label{ack}
 This work is supported by the EU Collaborative Project QuProCS (Grant Agreement 641277).


\begin{thebibliography}{99}
\bibitem{nota-fractional} If all the system-field stationary states are unbounded, the excitation of an initially excited atom eventually spreads over all space 
and the atom decays to its ground state. This is not the case if there is a bound atom-photon state: the atom's fraction of excitation overlapping this state will remain confined next to the atom, hence this will not decay fully (fractional decay).
\bibitem{nota-BCs} Since we are interested in the emission process into an infinite-length CCA, one can conveniently assume cyclic boundary conditions.
\bibitem{nota-T} Unlike \rref\cite{scirep} which was limited to the weak-coupling approximation, here the initial state $|0\rangle|{\rm vac}\rangle$ can overlap atom-photon bound states \cite{lambro} yielding that the atomic decay is generally {\it fractional}.
\end{thebibliography}
\end{document}